\documentclass[showpacs,prl,twocolumn,aps]{revtex4}

\usepackage{amsmath}
\usepackage{amssymb}
\usepackage{latexsym}
\usepackage{amsfonts}
\usepackage{epsfig}
\usepackage{psfrag}
\usepackage{graphicx}

\newcommand{\ket}[1]{\left|#1\right>}

\newcommand{\bea}{\begin{eqnarray}}
\newcommand{\ea}{\end{eqnarray}}
\newcommand{\eea}{\end{eqnarray}}
\newcommand{\ord}{\,{\cal O}}

\begin{document}

\title{Quantum Zeno suppression of three-body losses in 
Bose-Einstein condensates}  

\author{R. Sch\"{u}tzhold and G. Gnanapragasam}

\affiliation{Fakult\"{a}t f\"{u}r Physik, Universit\"at Duisburg-Essen, 
D-47048 Duisburg, Germany}

\begin{abstract}
We study the possibility of suppressing three-body losses in 
atomic Bose-Einstein condensates via the quantum Zeno effect, 
which means the delay of quantum evolution by frequent 
measurements. 
It turns out that this requires very fast measurements with the rate 
being determined by the spatial structure of the three-body form factor, 
i.e., the point interaction approximation 
$\delta^3(\mathbf{r}-\mathbf{r'})$ is not adequate. 
Since the molecular binding energy $E_b$ provides a natural limit for 
the measurement rate, this suppression mechanism can only work if the 
form factor possesses certain special properties. 
\end{abstract}


\maketitle

To the best of our knowledge, Bose-Einstein condensates of trapped 
dilute alkali gases are  the coldest systems in the Universe \cite{cornell}.
Apart from their low temperature, atomic Bose-Einstein condensates
are an ideal laboratory for studying many-body quantum effects 
\cite{wket,angket}
because they can be well manipulated and controlled experimentally 
by using lasers and magnetic fields \cite{cw}.
However, since the gaseous state is not the real ground state 
of these atoms, inelastic collisions due to three-body recombination 
occur \cite{macek}.
In such an exothermic process,  three atoms collide to form a dimer 
and an unbound atom, which carries away the excess energy and momentum. 
The binding energy thus released is converted into the kinetic energies 
of the colliding partners (molecule and third atom) which leave the trap. 
As a result of this process, the condensate is susceptible to loss and 
heating \cite{heating,grimm}, thereby inevitably limiting its lifetime 
and the minimum achievable temperature. 
Consequently, one is lead to the question as to how one could suppress 
this undesired loss mechanism. 

In this work we address this important question using the idea of 
quantum Zeno effect which is basically the delay or even inhibition of 
the time evolution of a quantum state by performing frequent 
measurements \cite{m&s, itano, raizen}.
As an example, consider a simple two-level system undergoing 
Rabi oscillations of frequency $\omega$. 
When the initial state $|\psi_1\rangle$ gets rotated under the 
action of the total Hamiltonian $\hat H$, the state evolves to 
\begin{equation}
\label{discrete}
|\psi(t)\rangle = e^{-i\hat Ht}|\psi_1\rangle = 
\cos\omega t |\psi_1\rangle - i \sin\omega t|\psi_2\rangle.
\end{equation}
Without measurements, the probability to find the state in 
$|\psi_1\rangle$ after time $t$ is 
$P(t) = |\langle\psi_1|e^{-iHt}|\psi_1\rangle|^2=\cos^2\omega t$. 
When $N$ measurements are made during this time
(after equidistant time intervals), the survival 
probability reads 
\begin{equation} 
\label{exp}
P_{N}(t) = \left[P\left(t/N\right)\right]^N 
\approx \exp\left(-\frac{t^2}{N\tau^2}\right),
\end{equation}
where $1/\tau^{2}=\langle\psi_1|\hat H^2|\psi_1\rangle=\omega^2$
\cite{facchi1}. 
For large $N$, the survival probability $P_{N}(t)$ approaches one,  
i.e., under continuous measurements the time evolution 
of the system is prevented.
Thus, quoting \cite{facchi1} 
\emph{Zeno's quantum-mechanical arrow (the wave function), 
sped by the Hamiltonian, does not move, if it is continuously observed}.

The above example demonstrates the suppression of transitions between 
two discrete levels in a quantum system.
However, the aforementioned three-body losses correspond to a decay 
into a continuum of states.
Thus, we have to study whether and when the quantum Zeno effect
occurs in such cases.
To this end, we consider \cite{facchi} the following model Hamiltonian 
$(\hbar=1)$ 
\bea
\label{ham}
\hat{H} = E_0|g\rangle \langle g| +
\int dk \left[E_k|k\rangle \langle k| + 
(g_k|g\rangle \langle k| + \mbox{h.c.})\right], 
\ea
where $\langle g|g\rangle =1$, $\langle k|k'\rangle = \delta(k-k')$, 
and $\langle g|k'\rangle = 0$.
Here $|g\rangle$ denotes an initial quasi-bound state 
(i.e., a resonance) with energy $E_0$ coupled to a continuum of 
states $|k\rangle$ with energies $E_k$ via the coupling strength $g_k$. 
Similar Hamiltonians have been used to study, for example, 
a two-level atom interacting with radiation field \cite{frasca}.

To the lowest order in perturbation $g_k$, the total decay probability 
$P_d(t)=\int dk |\langle k|e^{-i\hat{H}t}|g\rangle|^2$, takes the form
\begin{equation}
P_d(t) = 
t^2 \int dE\;|g(E)|^2\;\frac{\sin^2[(E-E_0)t/2]^2}{[(E-E_0)t/2]^2},
\label{rose}
\end{equation}
where we have used $|g(E)|^2 = |g_k|^2(dE_k/dk)^{-1}$. 
For short times, we obtain a quadratic growth of the decay probability
$P_d(t) \propto t^2$ similar to the previous example in 
Eq.~(\ref{discrete}).
Consequently, the time evolution can be delayed via the quantum 
Zeno effect according to Eq.~(\ref{exp}).
For longer times (but still within the region of perturbation theory),  
the decay probability grows linearly $P_d(t) \propto t$ and we 
recover Fermi's golden rule.  
In this regime, measurements would not slow down the time evolution 
significantly. 
Therefore, the crossover time $T_*$ when the transition between these 
two different regimes takes place \cite{T_*}, i.e.,  
$P_d(t\ll T_*) \propto t^2$ and $P_d(t\gg T_*) \propto t$, 
determines how fast one has to measure in order to slow down the 
decay via the quantum Zeno effect. 
We may estimate the crossover time $T_*$ via calculating the 
intersection point of the two asymptotic expressions 
$P_d(t\ll T_*) \propto t^2$ and $P_d(t\gg T_*) \propto t$ 
which yields 
\begin{equation}
\frac{1}{T_*} 
= 
\mathcal{O}\left(\int dE\frac{|g(E)|^2}{|g(E_0)|^2}\right) 
= 
\mathcal{O}(\Delta E).
\label{ct1}
\end{equation}
This is reminiscent of Heisenberg's energy--time uncertainty relation
and implies that one has to measure before the energy conservation 
begins to play a role in order to observe the quantum Zeno effect. 

Alternatively, using the wave-packet description one may write 
$T_*$ in the following way
\begin{equation}
\frac{1}{T_*} = \mathcal{O}(v_{\rm{group}}\Delta k) = 
\mathcal{O}\left(\frac{v_{\rm{group}}}{\Delta r}\right).
\label{ct2}
\end{equation}
Here $\Delta k$ and $\Delta E$ are the widths of $|g_k|^2$ and 
$|g(E)|^2$, respectively, and $v_{\rm{group}} = dE_k/dk$ 
is the group velocity 
(such that $v_{\rm{group}}\Delta k \approx \Delta E$). 
For early times, the wave-packet created by the decay of the 
initial quasi-bound state $|g\rangle$ into the continuum of 
states $|k\rangle$ possesses the width $\Delta k$ in momentum 
space and thus it spreads over $\Delta r=\ord(1/\Delta k)$
in position space.
Thus, $T_*$ is the period during which this wave-packet of spatial 
size $\Delta r$ needs to move away with its group velocity 
$v_{\rm{group}}$. 
In other words, as long as the wave-packet is close to its 
initial position (and thus has large overlap with
the original quasi-bound state $|g\rangle$), its decay can be 
suppressed via the quantum Zeno effect -- once the wave-packet 
has crossed its own dimensions and moved away, measuring does 
not change much anymore.

\begin{figure}[ht!]
\includegraphics[height=4cm]{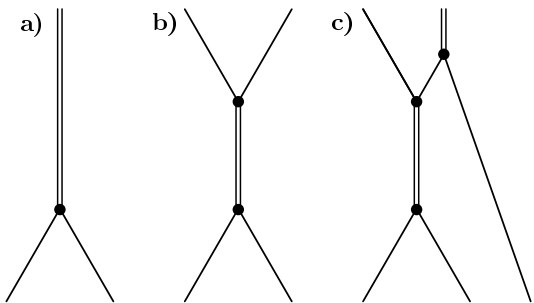}
\caption{Feynman diagrams for the Hamiltonian in  Eq.~(\ref{h}).
Single lines denote atoms and double lines are molecules while 
the black dot is the interaction vertex.
The left diagram a) shows the first-order process of direct molecule
formation which is forbidden by energy conservation. 
The middle diagram b) corresponds to atom-atom scattering via an 
intermediate (``virtual'') molecular state, which is a second-order 
effect.  
Finally, the right diagram c) combines the processes a) and b) and 
shows the actual third-order three-body recombination process 
resulting in the formation of a real molecule plus an atom 
(carrying away the excess energy and momentum). 
\label{fig1}}
\end{figure}

\begin{widetext}

Studying the simple toy model in Eq.~(\ref{ham}), we found that 
the quantum Zeno effect may also occur in the decay to a continuum
of states, but only if measurements are performed fast enough -- 
with the characteristic time $T_*$ being determined by the form 
factor $|g_k|^2$. 
Now let us study a more realistic model Hamiltonian for three-body 
losses in Bose-Einstein condensates.
For simplicity, we restrict our consideration to one species of atoms 
with mass $m$ (described by the field operator $\hat{\psi}$) coupled 
to one species of molecules with mass $2m$ 
(and the field operator $\hat{\phi}$)
\begin{eqnarray}
\hat{H} 
= 
\int d^3r  
\left\{
\hat{\psi}^{\dag}(\mathbf{r})
\left(\frac{-\nabla^2}{2m}\right)
\hat{\psi}(\mathbf{r})
+ 
\hat{\phi}^{\dag}(\mathbf{r})
\left(\frac{-\nabla^2}{4m}+E_b\right)
\hat{\phi}(\mathbf{r})
+ 
\int d^3r' 
\left[W(\mathbf{r}-\mathbf{r'})\hat{\phi}^{\dag}
\left(\frac{\mathbf{r}+\mathbf{r'}}{2}\right)
\hat{\psi}(\mathbf{r'})\hat{\psi}(\mathbf{r}) 
+ \mbox{h.c.}\right]
\right\}. 
\label{h}
\end{eqnarray}
Here $E_b$ is the negative molecular binding energy and 
$W(\mathbf{r}-\mathbf{r}')$ is the atom-molecule coupling. 
Since the characteristic length scales of molecule formation 
are typically much smaller than the trap size, we assume a 
homogeneous condensate with density $\rho$
and omit the trap potential $V(\mathbf{r})$.
Within standard perturbation theory (i.e., expansion into powers 
of the coupling strength $W$) we recover the usual Feynman diagrams 
depicted in Fig.~\ref{fig1}.
Three-body losses are described by processes of third (or higher) 
order (i.e., Fig.~\ref{fig1}c) and the decay probability reads to 
lowest order in $W$
\begin{eqnarray}
P_d(t) 
=  
\rho^3 |\tilde{W}(0)|^2 
\int d^3k\,|\tilde{W}(\mathbf{k})|^2 |\tilde{W}(2\mathbf{k})|^2 
\left|\frac{3}{4E_b}
\frac{e^{i\omega_1t}-1}{\omega_1\omega_3}
+\frac{1}{\omega_3-\omega_2}
\left(\frac{3}{4}\frac{e^{i\omega_2t}-1}{\omega_2\omega_3}-
\frac{e^{i\omega_3t}-1}{E_b\omega_3}
\right) \right|^2
\,.
\label{pink}
\end{eqnarray}
Here $\tilde{W}(\mathbf{k})=\tilde{W}(k)$ is the Fourier transform of 
$W(\mathbf{r}-\mathbf{r}')$ and we have abbreviated 
$\omega_1= E_b+3E_k/2$, 
$\omega_2 = E_b-E_k/2$,  
$\omega_3 = 3E_k/2$, and 
$E_k = k^2/(2m)$. 
Initially, at time $t=0$, we consider a situation 
$\ket{\Psi(t=0)}=\ket{\Psi_0}$ where all the 
atoms have Bose condensed 
$\hat\psi\ket{\Psi_0}=\sqrt{\rho}\ket{\Psi_0}$ 
and there are no molecules $\hat\phi\ket{\Psi_0}=0$. 
Of course, this is a strong simplification
(assuming perfect measurements and weak depletion etc.) 
but the main results should remain unaffected. 

\end{widetext}

The expression for the time dependence of the decay probability in 
Eq.~(\ref{pink}) allows us to study the occurrence of the quantum 
Zeno effect. 
For small $t$, a Taylor expansion of the integrand in 
Eq.~(\ref{pink}) in powers of $t$ yields 
\begin{eqnarray}
\label{t^6}
P_d(t) = 
\left(\frac{\rho^3 |\tilde{W}(0)|^2 }{36} 
\int d^3k\, |\tilde{W}(\mathbf{k})|^2 |\tilde{W}(2\mathbf{k})|^2\right)t^6
\,.
\end{eqnarray}
Thus, the decay probability increases as $t^6$ initially.
This departure from the conventional $t^2$ behavior obtained earlier 
is attributed to the three-body loss as a third-order process.
In this regime, frequent measurements slow down the decay significantly. 

For large $t$, on the other hand, we recover the linear time dependence 
analogous to Fermi's golden rule:
In this limit, the points of stationary phase in the integrand dominate 
the integral in Eq.~(\ref{pink}).
Since the binding energy $E_b$ is negative, we have $\omega_2<0$, 
$\omega_3>0$ and thus $\omega_3>\omega_2$ for all $k>0$. 
Consequently, the only point of stationary phase 
(which is also the only point where one of the denominators vanishes) 
is located at $\omega_1=0$ and (as one would expect) corresponds to 
energy conservation $3E_k/2=-E_b$. 
Thus we derive the asymptotic (large $t$) behavior of Eq.~(\ref{pink}) 
in complete analogy to Eq.~(\ref{rose}):
For large values of $t$, only the term containing 
$\sin^2(\omega_1t/2)/\omega_1^2$ exhibits a peak at 
$\omega_1=0$, of height $\propto t^2$ and width $\propto 1/t$.
Thus the significant contribution to the integral 
arises from the neighborhood of this peak. 
Performing the integration we get (for large $t$)
\bea
\label{t^1}
P_d(t) = 
\frac{256\pi^2\rho^3m^5}{27k_0^7}
\left|\tilde{W}(0)\tilde{W}(k_0)\tilde{W}(2k_0)\right|^2
\,t\,,
\ea
where the value of $k_0=(4mE_b/3)^{1/2}$ is dictated by 
energy conservation.

The crossover time $T_*$ between the two regimes (\ref{t^6}) and 
(\ref{t^1}) can again be estimated by equating these two asymptotic 
expressions for $P_d(t)$ 
\begin{equation}
\frac{1}{T_*^5} = 
\mathcal{O}\left(
\frac{3k_0^7}{1024\pi^2m^5}
\int d^3k
\left|
\frac{\tilde{W}(\mathbf{k})\tilde{W}(2\mathbf{k})}
{\tilde{W}(k_0)\tilde{W}(2k_0)}
\right|^2
\right) 
\,.
\label{ct}
\end{equation}
The integral in the equation above scales with the cubed width 
$k^3_W$ of the function $\tilde{W}(\mathbf{k})$ in $k$-space 
(relative to its value at $k_0$ and $2k_0$, 
respectively). 
Thus, the crossover time $T_*$ can be estimated as
\begin{equation}
\frac{1}{T_*^5} =
\mathcal{O}\left(\frac{ k^3_Wk_0^7}{m^5}\right) = 
\mathcal{O}\left(E_b^5\frac{k^3_W}{k_0^3}\right).
\label{ct3}
\end{equation}
Similar to the previous example in Eq.~(\ref{ham}), this expression 
facilitates an intuitive interpretation -- even though it is more 
qualitative than the quantitative picture described after 
Eq.~(\ref{ct2}).
At early times, the spatial size $\Delta r$ of the molecular 
wave-packet scales with $\Delta r=\ord(1/k_W)$ while its group 
velocity is determined by $k_0$, i.e., the binding energy $E_b$.
Hence $T_*$ grows for both $k_W$ and/or $k_0$ shrinking. 

\begin{figure}[ht!]
\includegraphics[height=3.8cm]{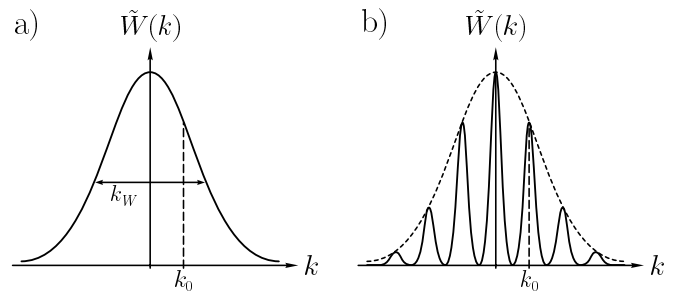}
\caption{Sketch of two possible forms for the atom-molecule coupling 
$\tilde{W}$ as a function of $k$ illustrating the two characteristic cases:
In the left picture a), $\tilde{W}$ has no sub-structure and thus 
$k_0$ and $k_W$ are of the same order of magnitude $k_W\sim k_0$,
i.e., the quantum Zeno effect is not applicable. 
(Note that $k_W\gg k_0$ would still be possible in this case -- 
but in this limit, the quantum Zeno effect does also not apply.)
In the right picture b), the function $\tilde{W}$ has more structure 
and thus we have $k_W\ll k_0$, which enables us to apply the quantum 
Zeno effect. 
\label{fig2}}
\end{figure}

In units of the inverse binding energy $E_b$, the crossover time $T_*$
scales with the ratio $k_0/k_W$ to the power $3/5$. 
In order to achieve a significant delay of the time evolution via the 
quantum Zeno effect, we would have to measure, i.e., detect the formed 
molecule, faster than this time scale $T_*$.
The presence of the molecule could be detected via driving a light 
induced transition between molecular bound states which is off-resonant 
for all atomic states (see also \cite{search}).
Then the energy-time uncertainty relation implies that this measurement 
cannot be done faster than the inverse binding energy $E_b$. 
Even if we could imagine a detection method for which this energy-time 
uncertainty bound does not apply, it would be unwise to measure 
faster than the inverse binding energy $E_b$: 
For such measurements, the direct two-body recombination process in 
Fig.~\ref{fig1}a which is forbidden by energy conservation, would start 
to contribute.
Since the rate of this first-order process (if allowed) is typically 
much larger than that of the third-order effect in Fig.~\ref{fig1}c,
the condensate would become even more unstable. 
Therefore, stabilizing the Bose-Einstein condensate via the quantum 
Zeno effect is only possible if $T_*E_b\gg1$.
With Eq.~(\ref{ct3}) we see that this requires $k_0\gg k_W$. 
So if the structure of $\tilde{W}(\mathbf{k})$ is basically 
determined by just one length scale 
(e.g., the $s$-wave scattering length $a_s$) 
as in Fig.~\ref{fig2}a, we cannot stabilize the Bose-Einstein 
condensate via the quantum Zeno effect.
However, if $\tilde{W}(\mathbf{k})$ has a more complicated, 
e.g., spiky, structure as in Fig.~\ref{fig2}b, we could have 
$k_0\gg k_W$ and thus it could be possible. 

In summary, we studied the possibility of suppressing three-body 
recombinations in dilute atomic Bose-Einstein condensates via the 
quantum Zeno effect.
This fascinating phenomenon corresponds to the delay or even inhibition 
of quantum evolution by repeated measurements if they are fast enough. 
Addressing the question of how fast is fast enough, we first see that 
the probability for a positive measurement outcome (saying that the 
system has actually evolved) must always be small. 
In addition to this trivial bound, another time-scale arises in the 
decay of a meta-stable state into a continuum:
In such a set-up, we obtain a quadratic (or higher polynomial) growth 
of the decay probability at early times (``quantum Zeno regime'') 
and a linearly increasing  decay probability at later times 
(``Fermi golden rule regime'').
Obviously, measurements can only slow down the evolution during the 
initial period (i.e., the ``quantum Zeno regime'').
For the model Hamiltonian in Eq.~(\ref{ham}), describing the tunnelling 
of a particle out of a potential well, for example, the crossover time 
$T_*$ separating the two regimes \cite{T_*}
is given by the width of the corresponding 
wave-packet in comparison with its group velocity. 
In the case of  three-body recombinations in dilute atomic Bose-Einstein 
condensates, this crossover time $T_*$ is determined by the 
spatial structure of the three-body form factor 
$W(\mathbf{r}-\mathbf{r'})$ in Eq.~(\ref{h}). 
As a result, the often used \cite{timm,holland} point interaction 
approximation 
$W(\mathbf{r}-\mathbf{r'})\propto\delta^3(\mathbf{r}-\mathbf{r'})$ 
is not adequate for this question. 
Since the molecular binding energy $E_b$ provides a natural limit for 
the measurement rate, this suppression mechanism can only work if the 
typical width $k_W$ of the form factor $\tilde W$ is much smaller 
than characteristic momentum of the final molecule/atom 
$k_0=(4mE_b/3)^{1/2}$ \cite{loosely}. 

It might be illustrative to compare our approach with the work 
\cite{search} in which the inhibition of three-body recombination 
via resonant $2\pi$ laser pulses was studied.
First of all, it should be noted that these $2\pi$ laser pulses 
considered in \cite{search} do not correspond to an actual measurement 
but induce a phase shift.
Therefore, the mechanism studied in \cite{search} is not the quantum 
Zeno effect, but corresponds to passive error correction techniques 
known in quantum information theory under the names 
``spin-echo'' or ``bang-bang'' method, see, e.g., \cite{bang}. 
Nevertheless, both phenomena are based on the crucial difference 
between summing amplitudes and probabilities in quantum theory.
Therefore, the repetition rate of the $2\pi$ laser pulses considered 
in \cite{search} should lie in the ``quantum Zeno regime'' in order 
to induce a significant slow-down (in the ``Fermi golden rule regime''  
they would have very little effect), i.e., similar temporal constraints 
should apply in both cases. 

RS acknowledges support from DFG (SFB-TR12). 
GG is indebted to Prof. S. Thomae for general fruitful discussions, 
kind help with the figures and hospitality. 



\begin{thebibliography}{9999}
\bibitem{loosely}
It has been shown that the recombination into loosely bound molecules,
with the pseudopotential interaction and a positive $s$-wave scattering 
length $a_s$, obeys universal features in the sense that all relevant 
quantities are unique functions of $a_s$.
%
For example, the binding energy $E_b$ of the $s$-wave bound state takes 
the simple form $E_b = -1/(ma_s^2)$ \cite{stoof+goral} from which it 
follows that $k_0=2/\sqrt{3}a_s$. 
%
If this universality also extends to the internal structure of $W$, 
then the quantum Zeno effect is not applicable in this case.
\bibitem{T_*}
Note that the crossover time $T_*$ in Eqs.~(\ref{ct1}) and 
(\ref{ct}) should not be confused with the time scale $\tau$ 
in Eq.~(\ref{exp}) which is sometimes \cite{facchi1} called 
``quantum Zeno time''.
\end{thebibliography}
\end{document}